\documentclass{Rinton-P9x6}

\newcommand{\newc}{\newcommand} 
\newc{\beq}{\begin{equation}} 
\newc{\eeq}{\end{equation}} 
\newc{\barr}{\begin{eqnarray}} 
\newc{\earr}{\end{eqnarray}} 
\begin{document}

\title{SUSY Cold Dark Matter Detection at large $\tan\beta$}

\author{M. E. G\'omez}
\address{ Centro de F\'{\i}sica das 
Interac\c{c}\~{o}es Fundamentais (CFIF),  
Departamento de F\'{\i}sica, \\ Instituto Superior T\'{e}cnico, 
Av. Rovisco Pais, 1049-001 Lisboa, Portugal.\\ 
E-mail:mgomez@cfif.ist.utl.pt}

\author{  J. D. Vergados} 
\address{ Theoretical Physics Division, University of Ioannina, \\
E-mail: vergados@cc.uoi.gr}

\maketitle

\abstracts{We study the direct detection rate for SUSY cold dark m
atter (CDM) predicted
by the minimal supersymmetric standard model with universal boundary 
conditions and large values for $tan\beta$.
The relic abundance of the lightest supersymmetric particle (LSP),
assumed to be approximately a bino, is
obtained by including its coannihilations with the next-to-lightest 
supersymmetric particle (NLSP), which is the lightest s-tau. 
We find  detectable rates in the currently planned experiments for a
sector of the parameter space consistent with the
cosmological constraint on the LSP relic abundance and the ones 
imposed by $b\rightarrow s \gamma$ and the Higgs searches.
}
\section{Introduction}
\par
The minimal Supersymmetric extension of the standard model (MSSM) 
\cite{haber} with $R$--parity conservation predicts an stable 
lightest supersymmetric particle (LSP). Its relic abundance 
can provide the  desirable amount of cold dark matter
(CDM) in order to close the Universe \cite{jun}. 

In the present work will concentrate on the more restrictive version 
of the MSSM, with minimal supergravity (mSUGRA) and 
gauge unification. We will show how at large values of $\tan\beta$ 
it is possible to find scenarios such that predicted 
LSP is detectable in currently planned experiments and its relic 
abundance falls inside the bounds of cosmological interest. 

\section{The MSSM at large $\tan\beta$ and the $b\rightarrow s \gamma$ 
constraint}

Some details on the choice of the MSSM 
parameter space have  already been presented by J.D. Vergados in this
conference \cite{vtalk}. Further details are given in Refs.~\cite{cdm,nos}. 

The most two relevant characteristics of the MSSM at large 
$\tan\beta$ for our study are the relatively low values of the 
masses of pseudoscalar higgs $m_A$ and the NLSP (the 
lightest stau in our analysis). The fist is 
related to the enhancement of the LSP--nucleon 
scalar cross section at large $\tan\beta$, the second enables 
coannihilations LSP--NLSP which are required for the prediction of
a LSP relic abundance inside the cosmological bounds.

An accurate determination of $m_A$ is crucial for our work. We follow 
the procedure outlined in ref.~\cite{cdm}, which takes into account the full  
1--loop potential effective potential. Imposing a relation between the 
LSP and NLSP masses we can find the values of $m_0$ and $M_{1/2}$ 
corresponding to a certain value of $m_A$. Therefore the GUT 
values of $M_{1/2}$ and  $m_{0}$ can be  traded by the value of
 $m_A$ and the mass splitting between the
LSP and the NLSP $\Delta_{\tilde\tau_2}=(m_{\tilde\tau_2}-m_{\tilde\chi})/
m_{\tilde\chi}$.   

The values of our input parameters in the two scenarios we consider are
shown in fig.1. The higher one, 
$\tan\beta=52$, corresponds approximately to the unification of the 
tau and top Yukawa couplings at $M_{GUT}$. 
The lower value, $tan\beta=40$, results in a  $\sigma^{nucleon}_{scalar}$ 
smaller by one order of magnitude. $M_S$ is the common SUSY threshold, 
defined as $M_S=\sqrt{m_{\tilde t_1}m_{\tilde t_2}}$. The shaded 
areas correspond to the range of
values taken by $m_A$, $M_S$ as  $\Delta_{\tilde{\tau_2}}$ ranges from
0 to 1. The area associated with $m_0$ for the same range of 
$\Delta_{\tilde{\tau_2}}$ is wider as shown by the dashed and solid lines.

The choice  $\mu>0$ leads to a constraint on the parameter space arising
from the lower bound on $b\rightarrow s \gamma$. As a result the 
relatively light values for 
$m_\chi$ and the obtained detection rates are suppressed.
Our determination of $BR(b\rightarrow s \gamma)$ follows the procedure 
described in ref.~\cite{cdm2}.  We complete this analysis by including
the appropriate next-to-leading order (NLO) QCD corrections 
to the SUSY contribution 
at large values of $\tan\beta$ which recently have 
become available \cite{NLO}. The lower limits  
on $m_\chi$ resulting from this constraint are shown in fig.2.
\begin{figure}
\begin{minipage}[b]{9in}
\epsfig{figure=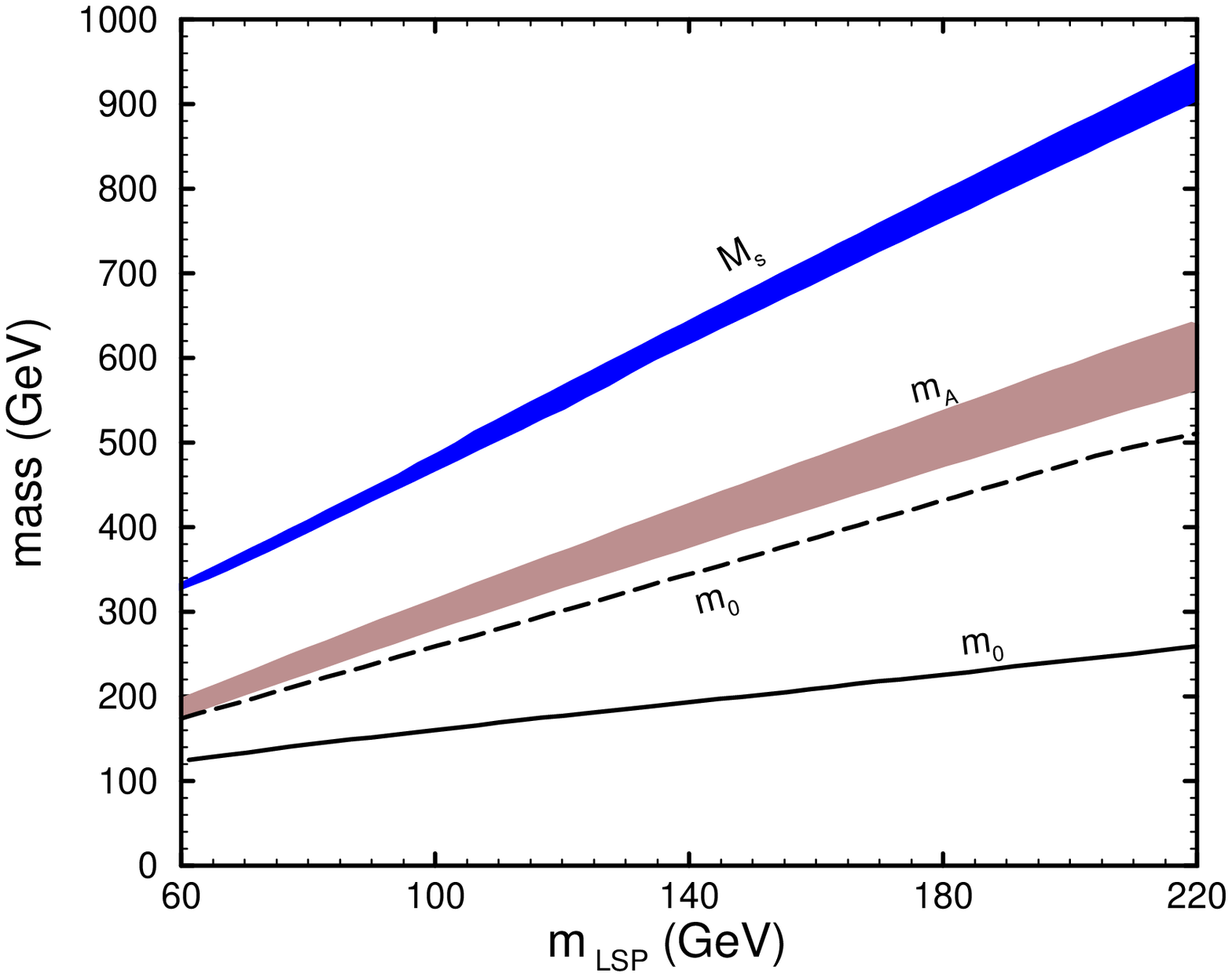,height=2.3in,width=2.3in,angle=0}
\epsfig{figure=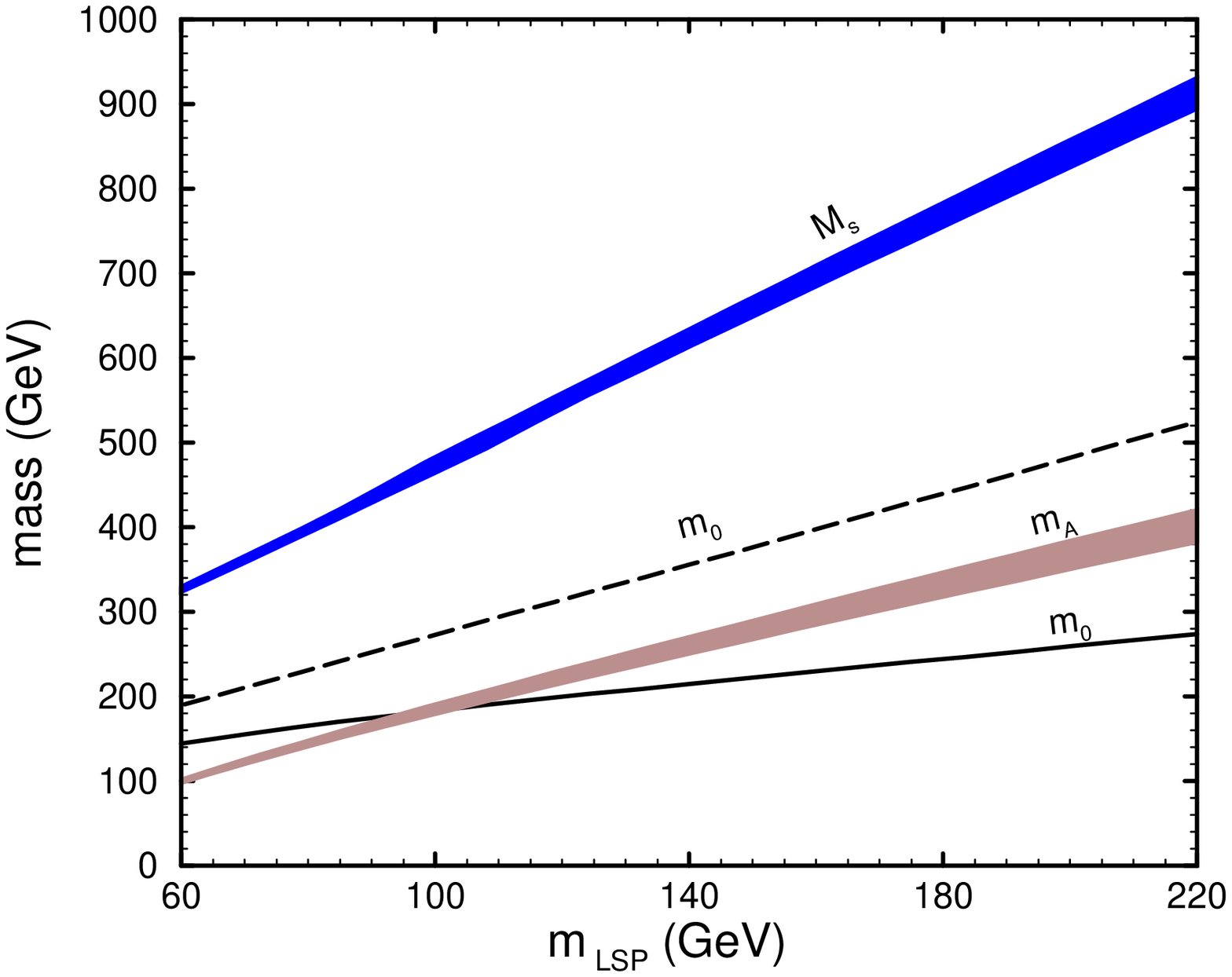,height=2.3in,width=2.3in,angle=0}
\end{minipage}
\medskip
\caption{The values of  $m_0$, $m_{A}$ 
and $M_S$ as functions of $m_{\tilde\chi}$ or $m_{\tilde{\chi}}$ for 
$\tan\beta=40$ (left) and $\tan\beta=52$ (right), 
$\mu>0$, $A_0=0$, the upper boundaries (lower) on the shaded areas
and the upper (lower) dashed line
corresponds to $m_{\tilde\tau_2}= 2 \times m_{\tilde\chi}$ 
($m_{\tilde\tau_2}=m_{\tilde\chi}$).
\label{maslsp}}
\end{figure}

\section{LSP relic abundance}

Following the considerations of Ref.~\cite{cdm} on the composition 
of the energy density of the universe in scenarios with vanishing 
and non vanishing cosmological constant we assume $\Omega_{LSP} h^2$ 
in the range:
\begin{equation}
0.09 \le \Omega_{LSP} h^2 \le 0.22
\label{eq:in2}
\end{equation}

The composition of the LSP on the model under consideration can be written
in the basis of the gauge and Higgs bosons superpartners as:
\begin{equation}
\tilde{\chi}\equiv\tilde{\chi}^0=C_{11}\tilde{B}+C_{12}\tilde{W}+
C_{13}\tilde{H}_1+C_{14}\tilde{H}_2.
\end{equation}
In the parameter space we study, $\tilde{\chi}$
is mostly a gaugino 
with $P=|C_{11}|^2+|C_{12}|^2 > .95$, with the Bino component being the most 
dominant one.

The fact that ($\tilde\chi$) is mostly a 
$\tilde B$ implies that the main contribution 
to its annihilation cross section arises from s-fermion 
(squark, s-lepton) exchange in the t- and u-channel 
leading to $f\bar f$ final states ($f$ is a quark 
or lepton). If, however, the mass of $\tilde\chi$ is close to the one 
of the NLSP, coannihilations between the two particles must be 
taken into account \cite{coan}. The inclusion 
these coannihilation effects results in a dramatic reduction of the 
($\tilde\chi$) relic abundance as the two lightest SUSY particles
approach in mass \cite{cdm,drees,ellis}. 
We estimate the  relic abundance of the LSP ($\tilde\chi$), 
by employing the analysis of Ref.\cite{cdm} which is appropriate for 
large $\tan\beta$ and includes coannihilations $\tilde\chi-\tilde{\tau}$,
suitable for Bino like LSP. To the list of coannihilation channels
is given in Table I.

\begin{center}
\vspace{1cm}
TABLE I. Feynman Diagrams 
\end{center} 
\begin{center}
\begin{tabular}{|c|c|c|} 
\cline{1-1}\cline{2-2}\cline{3-3} 
\multicolumn{1}{|c|}{Initial State} &
\multicolumn{1}{|c|}{Final State} &
\multicolumn{1}{|c|}{Diagrams}\\
\cline{1-1}\cline{2-2}\cline{3-3}
$\tilde\chi\tilde\chi$ & $\tau\bar\tau$ & 
$t(\tilde\tau_{1,2}),~u(\tilde\tau_{1,2})$
\\
& $e\bar e$ & $t(\tilde e_R),~u(\tilde e_R)$
\\ 
\hline  
$\tilde\chi\tilde\tau_2$ & $\tau h,~\tau H,
~\tau Z$ & 
$s(\tau),~t(\tilde\tau_{1,2})$
\\
& $\tau A$ & $s(\tau),~t(\tilde\tau_1)$
\\
& $\tau\gamma$ & 
$s(\tau),~t(\tilde\tau_2)$
\\ 
\hline
$\tilde\tau_2\tilde\tau_2$ & $\tau\tau$ & 
$t(\tilde\chi),~u(\tilde\chi)$
\\ 
\hline
$\tilde\tau_2\tilde\tau_2^\ast$ & 
$~hh,~hH,~HH,~ZZ~$ & $~s(h),~s(H),
~t(\tilde\tau_{1,2}),~u(\tilde\tau_{1,2}),~c~$
\\
& $AA$ & $~s(h),~s(H),~t(\tilde\tau_1),
~u(\tilde\tau_1),~c$
\\   
& $H^+ H^-,~W^+ W^-$ & $s(h),~s(H),~s(\gamma),~s(Z),~c,~t(\tilde\nu_\tau)$
\\ 
& $\gamma\gamma,~\gamma Z$ & $t(\tilde\tau_2),
~u(\tilde\tau_2),~c$
\\   
& $t\bar t,~b\bar b$ &
$s(h),~s(H),~s(\gamma),~s(Z)$
\\   
& $\tau\bar\tau$ & $s(h),~s(H),~s(\gamma),~s(Z),
~t(\tilde\chi)$    
\\   
& $u\bar u,~d\bar d,~e \bar e$ & $s(\gamma),~s(Z)$
\\   
\cline{1-1}\cline{2-2}\cline{3-3}                  
\end{tabular}
\vspace{1cm}
\end{center}

We should, at this point, clarify that 
in the parameter space considered
here no resonances in the s--channels were found. In other words
the s--channel exchange of 
A , h, H, Z into  $\tilde{\tau_2} \tilde{\tau_2}^*$ never becomes
resonant in the parameter space of our analysis. We can see in Fig. 1, however,
that a line $mass= 2 m_{\tilde{\chi}}$ will be above of the $m_A$ region 
for the 
case of $\tan\beta=52$, while for $\tan\beta=40$ it will not. 
However we should emphasize here, that the position the $m_A$ band 
displayed is Fig.1 respect a line of $mass= 2 m_{\tilde{\chi}}$ is 
very sensitive to small changes in $\tan\beta$ and  
the values $m_t$,$m_b$ and the GUT values for  $A_0$ and $m_0$. 
Therefore, at the large values of $\tan\beta$ it is possible 
to find sectors of the space of parameters where 
$m_A\approx m_{\tilde{\chi}}$, in these cases the the adequate treatment
of the Higgs mediated annihilation channels will be determining for an accurate
calculation of $\Omega_{LSP}~h^2$. 

The choice of parameter space in the two examples we present is aimed
to illustrate the decisive role of $\tan\beta$ in the LSP 
detection rates as we show in Fig.2. In the two scenarios 
we choose, annihilation resonant 
channels are not present and coannihilations are required in order 
to predict a cosmologically desirable LSP relic abundance. 

\begin{figure}
\begin{minipage}[b]{9in}
\epsfig{figure=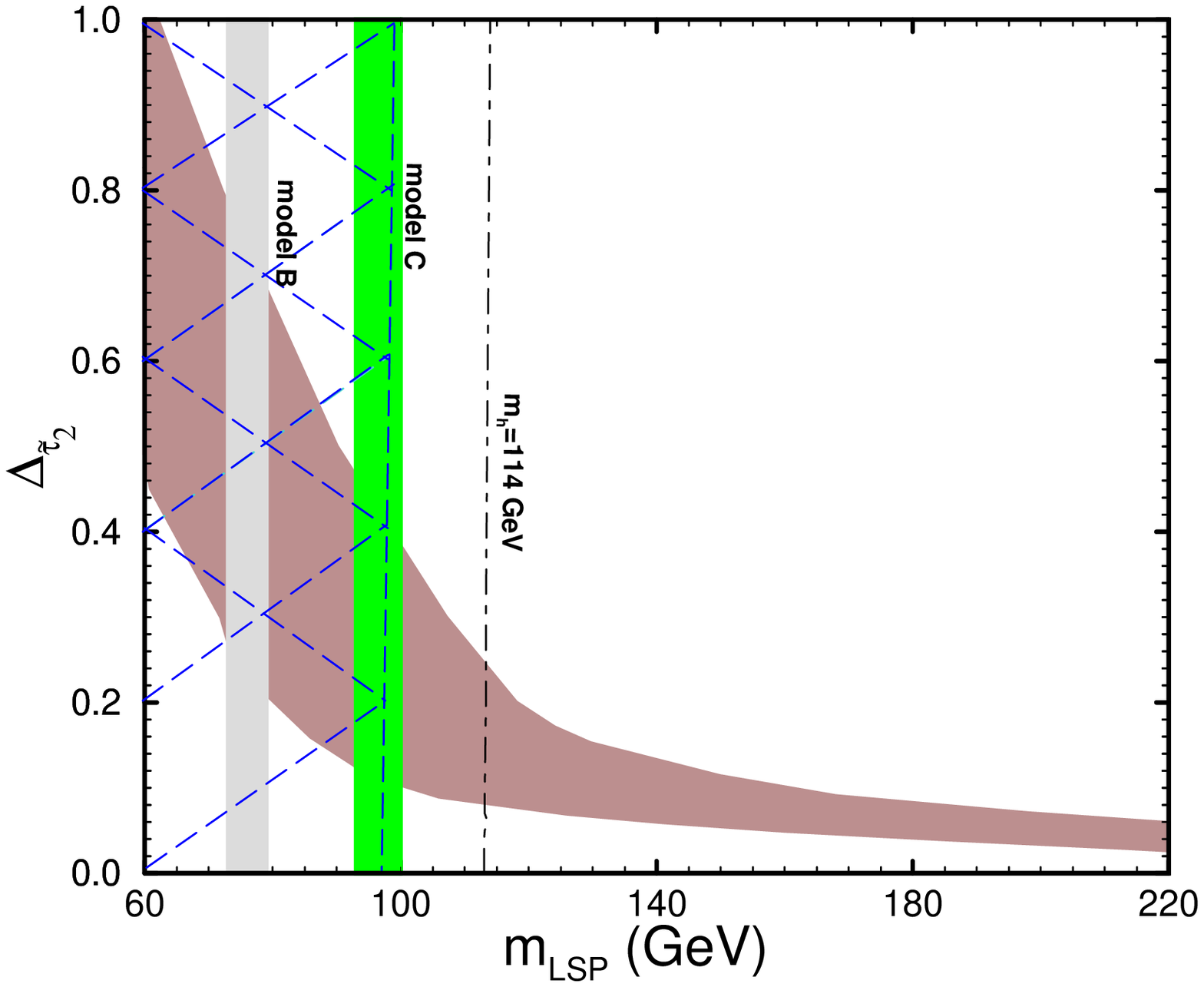,height=2.5in,width=2.3in,angle=0}
\epsfig{figure=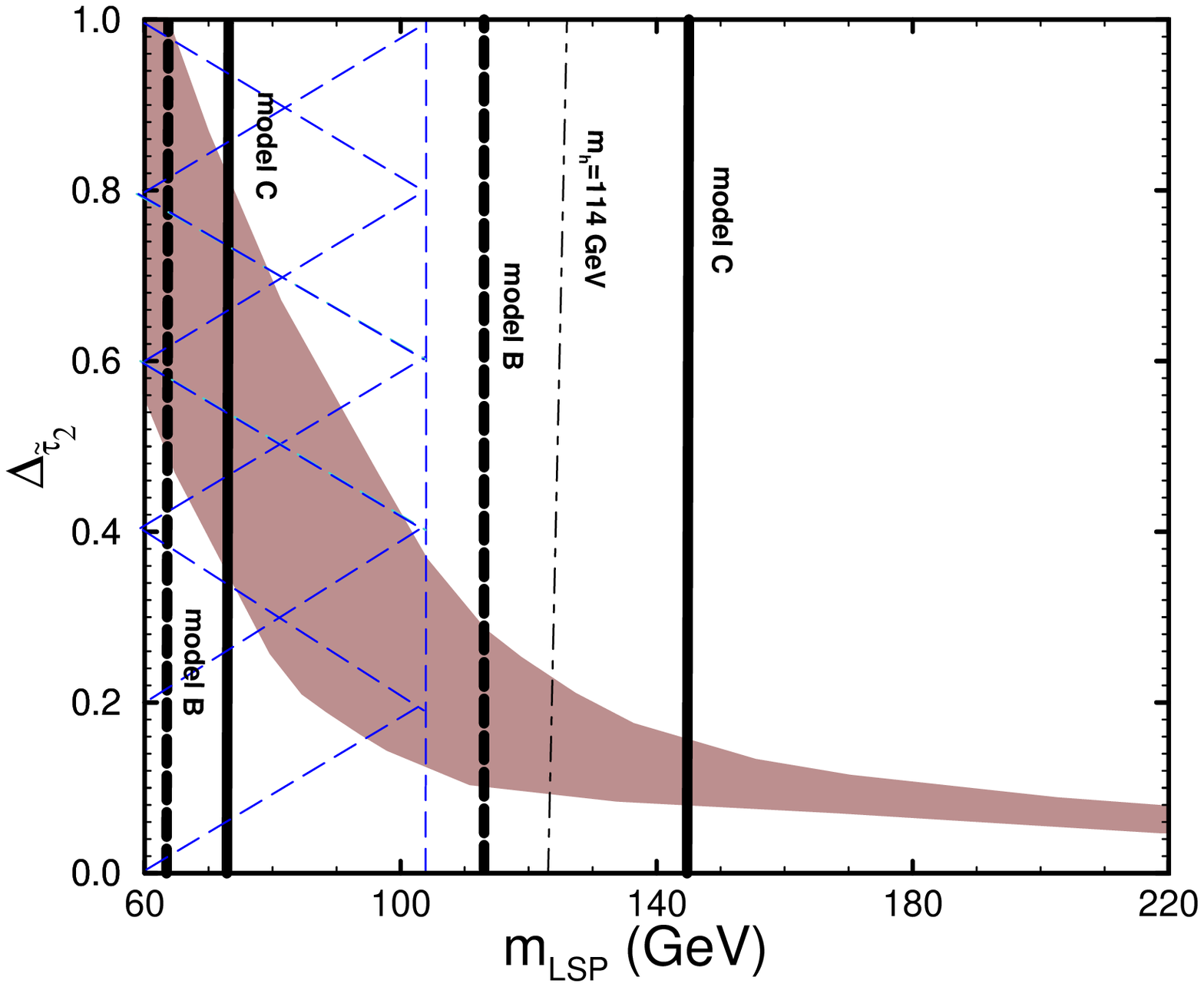,height=2.5in,width=2.3in,angle=0}
\end{minipage}
\medskip
\caption{The cosmologically allowed region in the 
$m_{\tilde{\chi}}-\Delta_{\tilde\tau_2}$ plane (shaded area) for $\tan\beta=40$ (left)
and $\tan\beta=52$ (right). 
The vertical bands in graph on  the left
correspond to the bound $\sigma_{scalar}^{(nucleon)}=4 \cdot10^{-7} pb$, 
obtained
from figure 2 for both models.
The vertical lines on the graph of the right 
correspond to the bounds $\sigma_{scalar}^{(nucleon)}=4 \cdot10^{-7} pb$
(lines towards the right of the graph) and $2 \cdot 10^{-5} pb$ for 
the models indicated. The marked areas on the left are excluded by 
$b\rightarrow s \gamma$.
\label{area50}}
\end{figure}

\section{LSP--Nucleon Elastic Cross Section}

The coherent scattering 
$\tilde\chi \, +\, (A,Z) \, \to \, \tilde\chi \,  + \, (A,Z)^*$ 
can be mediated via s-quarks and neutral 
Higgs particles (h and H). In our model we find that the Higgs contribution 
becomes dominant and therefore:
\begin{equation}
\sigma^{nucleon}_{scalar}\propto\left[f_s^0-f_s^1(1-2\frac{Z}{A})\right]^2,
\end{equation}
where:
\begin{eqnarray}
f_s^0&=&\frac{1}{2}(g_u+g_d)+g_s+g_c+g_b+g_t\\
f_s^1&=&\frac{1}{2}(g_u-g_d).
\end{eqnarray}
With:
\begin{eqnarray}
g_{u_i}&=&\left[g_h cos\alpha +g_H sin\alpha\right]\frac{f_{u_i}}{sin\beta},
\ \ u_i=u,c,t;\\
g_{d_i}&=&\left[-g_h sin\alpha +g_H cos\alpha\right]\frac{f_{d_i}}{sin\beta},
\ \ d_i=d,s,b.
\end{eqnarray}

In the eq, above $\alpha$ is the mixing angle which appear in the 
diagonalizaton of the $CP$-even Higgs mass matrix, and $g_h,\ g_H$ can be 
written as:
\begin{eqnarray}
g_h&=& 4 (C_{11}^\star tan\theta_W-C_{11}^\star)
(C_{41}\cos\alpha+C_{31}\sin\alpha)\frac{m_N m_W}{m_h^2}\\
g_H&=& 4 (C_{11}^\star tan\theta_W-C_{11}^\star)
(C_{41}\sin\alpha-C_{31}\cos\alpha)\frac{m_N m_W}{m_H^2}.
\end{eqnarray}

The last equations  gets enhanced as  $\tan\beta$ increases, since  the 
electroweak symmetry breaking imposes lower values for the 
pseudoscalar Higgs mass $m_A$ (see Fig. 1). This implies a lower value 
for of $m_H$. The changes on   $m_h$ are not so important, 
since its value can only move below an upper bound of 
about 120-130 GeV .
The coefficients $C_{ij}$ depends on the composition of the LSP. 
For all the values of the $m_{\tilde{\chi}}$, however, reported 
in the present work the condition $ P>.9$ is maintained. It becomes even more
stringent, $P >.95$, for $m_{\tilde{\chi}}>100 \ \rm{GeV}$.

The factors $f_{u_i},\ f_{d_i}$ parametrize the quark nucleon matrix 
element. They depend on the quark model used for the nucleon, we use 
two different quark models in our calculation denoted as 
models B and C. 
Their values along with further details were given in by Vergados 
\cite{vtalk} and can be found in Ref.~\cite{nos}.

The parameter space leading to predictions of 

\begin{equation}
4\times 10^{-7}~pb~ \le \sigma^{nucleon}_{scalar} 
\le 2 \times 10^{-5}~pb~
\label{eq:in3}
\end{equation}

is shown in fig.2.

\section{Conclusions}

In summary, we have found that the most popular version of the MSSM with 
gauge unification and universal boundary conditions at the GUT scale,
and a parameter space determined by large values of $\tan\beta$,
can accommodate a cosmologically suitable LSP relic abundance and
predict detection rates, which can be tested in current or projected
experiments.

We should mention that the calculated detection rates can 
vary by orders of magnitude, depending
on the yet unknown LSP mass.  Other source of  uncertainty comes 
from estimating the heavy quark contribution in the
nucleon cross section. This seems to be under control. We take the difference
between the models B and C discussed above as an indication of such
uncertainties. They seem to imply uncertainties no more than factors of two.  

 We believe, therefore, that, concerning the direct LSP detection event rates 
the main uncertainties come from the fact that the SUSY parameter space 
is not yet sufficiently constrained. The parameter space may be sharpened
by the accelerator experiments, even if the LSP is not found.  We should
mention here, in particular, the Higgs searches, since, as we have seen, the 
role of the Higgs particles in direct SUSY dark matter detection is crucial. 
It is not an exaggeration to say that
the underground and accelerator experiments are complementary and should
achieve a symbiosis.

This work was supported by the European Union under the contracts 
RTN No HPRN-CT-2000-00148 and TMR 
No. ERBFMRX--CT96--0090 and $\Pi E N E \Delta~95$ of the Greek 
Secretariat for Research.

\def\ijmp#1#2#3{{ Int. Jour. Mod. Phys. }{\bf #1~}(#2)~#3}
\def\pl#1#2#3{{ Phys. Lett. }{\bf B#1~}(#2)~#3}
\def\zp#1#2#3{{ Z. Phys. }{\bf C#1~}(#2)~#3}
\def\prl#1#2#3{{ Phys. Rev. Lett. }{\bf #1~}(#2)~#3}
\def\rmp#1#2#3{{ Rev. Mod. Phys. }{\bf #1~}(#2)~#3}
\def\prep#1#2#3{{ Phys. Rep. }{\bf #1~}(#2)~#3}
\def\pr#1#2#3{{ Phys. Rev. }{\bf D#1~}(#2)~#3}
\def\np#1#2#3{{ Nucl. Phys. }{\bf B#1~}(#2)~#3}
\def\npps#1#2#3{{ Nucl. Phys. (Proc. Sup.) }{\bf B#1~}(#2)~#3}
\def\mpl#1#2#3{{ Mod. Phys. Lett. }{\bf #1~}(#2)~#3}
\def\arnps#1#2#3{{ Annu. Rev. Nucl. Part. Sci. }{\bf
#1~}(#2)~#3}
\def\sjnp#1#2#3{{ Sov. J. Nucl. Phys. }{\bf #1~}(#2)~#3}
\def\jetp#1#2#3{{ JETP Lett. }{\bf #1~}(#2)~#3}
\def\app#1#2#3{{ Acta Phys. Polon. }{\bf #1~}(#2)~#3}
\def\rnc#1#2#3{{ Riv. Nuovo Cim. }{\bf #1~}(#2)~#3}
\def\ap#1#2#3{{ Ann. Phys. }{\bf #1~}(#2)~#3}
\def\ptp#1#2#3{{ Prog. Theor. Phys. }{\bf #1~}(#2)~#3}
\def\plb#1#2#3{{ Phys. Lett. }{\bf#1B~}(#2)~#3}
\def\apjl#1#2#3{{ Astrophys. J. Lett. }{\bf #1~}(#2)~#3}
\def\n#1#2#3{{ Nature }{\bf #1~}(#2)~#3}
\def\apj#1#2#3{{ Astrophys. Journal }{\bf #1~}(#2)~#3}
\def\anj#1#2#3{{ Astron. J. }{\bf #1~}(#2)~#3}
\def\mnras#1#2#3{{ MNRAS }{\bf #1~}(#2)~#3}
\def\grg#1#2#3{{ Gen. Rel. Grav. }{\bf #1~}(#2)~#3}
\def\s#1#2#3{{ Science }{\bf #1~}(19#2)~#3}
\def\baas#1#2#3{{ Bull. Am. Astron. Soc. }{\bf #1~}(#2)~#3}
\def\ibid#1#2#3{{ ibid. }{\bf #1~}(19#2)~#3}
\def\cpc#1#2#3{{ Comput. Phys. Commun. }{\bf #1~}(#2)~#3}
\def\astp#1#2#3{{ Astropart. Phys. }{\bf #1~}(#2)~#3}
\def\epj#1#2#3{{ Eur. Phys. J. }{\bf C#1~}(#2)~#3}

\end{document}